~ ~ ~ ~ ~ ~ ~ ~ ~ ~ ~ ~

\documentstyle[preprint,aps]{revtex}
\begin{document}
\title{Wigner Distribution Analysis of a Schr\"{o}dinger Cat Superposition of
Displaced Equilibrium Coherent States}
\author{{G.W. Ford$\dagger$ and R.F. O'Connell$\ddagger$\thanks{%
Corresponding Author: \newline
{\em e-mail: rfoc@rouge.phys.lsu.edu} (R. F. O'Connell)}}}
\address{ School of Theoretical Physics, Dublin Institute for Advanced
Studies \\
10 Burlington Road, Dublin 4, Ireland}
\date{\today}
\maketitle

\begin{abstract}
Motivated by recent experiments, we consider a Schr\"{o}dinger cat
superposition of two widely separated coherent states in thermal
equilibrium. The time development of our system is obtained using Wigner
distribution functions.  In contrast to our discussion for a two-Gaussian wave
packet [Phys. Lett. A 286 (2001) 87], we find that, in the absence of
dissipation, the interference term does not decay rapidly in time, but in common
with the other two terms, it oscillates in time and persists for all times.
\end{abstract}


Decoherence phenomena are at the forefront of cutting-edge research in
fundamental and applied quantum mechanics, as summarized in \cite{ford02},
where we stressed that, at variance with common lore, decoherence without
dissipation could occur in the case of a prototypical Schr\"{o}dinger cat
state of two widely separated Gaussians. The question arises as to whether
this is a general phenomena so here, motivated by the NIST experiments on
trapped ions \cite{myatt00,turchette00}, we examine decoherence without
dissipation for the case of a widely separated pair of equilibrium coherent
states.

Consider {\underline{any}} quantum state consisting of two identical
components separated by a distance $d$. Then, it is not difficult to show
that the corresponding Wigner distribution function, $W^{(2)}(q,p)$ at time $%
t=0$, say, is given by%
\begin{equation}
W^{(2)}(q,p,0)=N_{0}\left\{ W\left( q+\frac{d}{2},p,0\right) +W\left( q-%
\frac{d}{2},p,0\right) +2\cos \left( \frac{pd}{\hbar }\right)
W(q,p,0)\right\} ,  \label{1cat}
\end{equation}%
where $W(q,p,0)$ is the Wigner function for one of the pairs at $t=0$ and $%
N_{0}$ is a normalization factor. We wish to examine the case where $%
W(q,p,0) $ is an {\underline{oscillator}} state and where $W(q\pm \frac{d}{2}%
,p,0)$ corresponds to the pair of coherent (displaced oscillator) states.

Decoherence is a measure of how the interference term in (\ref{1cat}) decays
in time relative to the other terms. Thus, we need to calculate $%
W^{(2)}(q,p,t)$. This calculation is facilitated by the fact that the
equation of motion for the Wigner function of an oscillator is the same as
the classical equation of motion \cite{hillery84}. Hence%
\begin{equation}
W(q,p,t)=W(q(-t),p(-t),0),  \label{2cat}
\end{equation}%
where%
\begin{equation}
q(t)=q\cos \omega t+\left( \frac{p}{m\omega }\right) \sin \omega t,
\label{3cat}
\end{equation}%
and%
\begin{equation}
p(t)=p\cos \omega t-m\omega q\sin \omega t,  \label{4cat}
\end{equation}%
where $\omega $ is the oscillator frequency. It follows that%
\begin{eqnarray}
W^{(2)}(q,p,t) &=&A\{W(q(-t)+\frac{d}{2},~p(-t),~0)+W(q(-t)-\frac{d}{2}%
,~p(-t),~0)  \nonumber \\
&&~~~{}+2\cos \left( \frac{p(-t)d}{\hbar }\right) W(q(-t),~p(-t),~0)\},
\label{5cat}
\end{eqnarray}%
and $A$ is now the appropriate normalization factor.

We now consider the case where the system is in thermal equilibrium. The
equilibrium Wigner function for the oscillator is \cite{hillery84}

\begin{equation}
W_{0}(q,p)=\frac{1}{2\pi m\sqrt{\langle \dot{q}^{2}\rangle \langle
q^{2}\rangle }}\exp \left\{ -\frac{p^{2}}{2m^{2}\langle \dot{q}^{2}\rangle }-%
\frac{q^{2}}{2\langle q^{2}\rangle }\right\} ,  \label{6cat}
\end{equation}%
where the subscript "$0$" indicates equilibrium and%
\begin{equation}
\langle \dot{q}^{2}\rangle =\omega ^{2}\langle q^{2}\rangle =\frac{\hbar
\omega }{2m}\coth \left( \frac{\hbar \omega }{2kT}\right) =\frac{\hbar
\omega }{2m}\left( 2N+1\right) .  \label{7cat}
\end{equation}%
It follows from (\ref{3cat}), (\ref{4cat}), and (\ref{6cat}) that%
\begin{equation}
W_{0}(q(-t),~p(-t),~0)=W_{0}(q,p,0).  \label{8cat}
\end{equation}%
Hence%
\begin{eqnarray}
W_{0}^{(2)}(q,p,t) &=&A_{0}W_{0}(q,p,0)(\exp \left\{ -\frac{1}{2\langle
q^{2}\rangle }\left[ \frac{d^{2}}{4}+d\left( q\cos \omega t-\frac{p}{m\omega 
}\sin \omega t\right) \right] \right\}   \nonumber \\
&&~~~~~~~~{}+\exp \left\{ -\frac{1}{2\langle q^{2}\rangle }\left[ \frac{d^{2}%
}{4}-d\left( q\cos \omega t-\frac{p}{m\omega }\sin \omega t\right) \right]
\right\}   \nonumber \\
&&~~~~~~~~{}+2\cos \left\{ \frac{(p\cos \omega t+m\omega q\sin \omega t)d}{%
\hbar }\right\} ),  \label{9cat}
\end{eqnarray}%
where%
\begin{equation}
A_{0}=\left( 2\left[ 1+\exp \left\{ -\frac{m^{2}\langle \dot{q}^{2}\rangle
d^{2}}{2\hbar ^{2}}\right\} \right] \right) ^{-1}.  \label{10cat}
\end{equation}

As before, decoherence is given a quantitative measure by defining an
attenuation coefficient $a_{w}(t)$, which is the ratio of the factor
multiplying the cosine to twice the geometric mean of the first two terms
(and here the subscript $w$ indicates decoherence in phase space to
distinguish it from the more physically meaningful quantity $a(t)$
corresponding to decoherence in coordinate space). Hence
\begin{equation}
a_{w}(t)=\exp \left\{ \frac{d^{2}}{8\langle q^{2}\rangle }\right\} .
\label{13cat}
\end{equation}

Next, we obtain the coordinate probability by integrating (\ref{9cat}) over $%
p$. First, we note, using (\ref{6cat}), that

\begin{eqnarray}
P_{0}(q,0) &=&\int_{-\infty }^{\infty }W_{0}(q,p,0)dp  \nonumber \\
&=&\left[ 2\pi \langle q^{2}\rangle \right] ^{-1/2}\exp \left\{
-\frac{q^{2}}{ 2\langle q^{2}\rangle }\right\} .  \label{14cat}
\end{eqnarray}%
It follows, from (\ref{9cat}) and (\ref{14cat}), that%
\begin{eqnarray}
P^{(2)}(q,t) &=&A_{0}P_{0}(q,t)(\exp \left\{ -\frac{1}{2\langle q^{2}\rangle 
}\left( \frac{d^{2}}{4}\cos ^{2}\omega t-qd\cos \omega t\right) \right\} 
\nonumber \\
&&~~~~~{}+\exp \left\{ -\frac{1}{2\langle q^{2}\rangle }\left( \frac{d^{2}}{4%
}\cos ^{2}\omega t+qd\cos \omega t\right) \right\}  \nonumber \\
&+&2\exp \left\{ -\frac{m^{2}\langle \dot{q}^{2}\rangle d^{2}}{2\hbar ^{2}}%
\cos ^{2}\omega t\right\} \cos \left( \frac{dm\omega q\sin \omega t}{\hbar }%
\right) ).  \label{15cat}
\end{eqnarray}%
Hence%
\begin{eqnarray}
a(t) &=&\exp \left\{ \frac{d^{2}}{8\langle q^{2}\rangle }\cos ^{2}\omega t-%
\frac{d^{2}m^{2}\langle \dot{q}^{2}\rangle }{2\hbar ^{2}}\cos ^{2}\omega
t\right\}  \nonumber \\
&=&\exp \left\{ -\frac{m\omega d^{2}\cos ^{2}\omega t}{2\hbar \sinh
\left( 
\frac{\hbar \omega }{kT}\right) }\right\} .  \label{16cat}
\end{eqnarray}

It is clear that we do not have an interference term which decays rapidly in
time but instead, in common with the other terms, it oscillates in time and
persists for all time. This is in contrast to what we found for a
two-Gaussian state of a free particle \cite{ford02,ford01}.  However,
for small times after the times for which the attenuation factor has
its maximum value of unity, and for negligibly small frequencies, we
obtain a decoherence decay time which is consistent with our results
for a free particle \cite{ford02,ford01}.  However, for small times
after the times for which the attenuation factor has its maximum
value of unity, and for negligibly small frequencies, we obtain a
decoherence decay time which is consistent with our results for a
free particle \cite{ford01}.

\end{document}